# Electrically injected GeSn lasers with peak wavelength up to 2.7 μm at 90 K

Yiyin Zhou,[1,2] Solomon Ojo,[1,2] Yuanhao Miao,[1] Huong Tran[1], Joshua M. Grant,[1,2] Grey Abernathy,[1,2] Sylvester Amoah,[1] Jake Bass,[1] Gregory Salamo,[3,4] Wei Du,[5] Jifeng Liu,[6] Joe Margetis,[7] John Tolle,[7] Yong-Hang Zhang,[7] Greg Sun,[8] Richard A. Soref,[8] Baohua Li,[9] Shui-Qing Yu[1,4†]

[1]Department of Electrical Engineering, University of Arkansas, Fayetteville, Arkansas 72701, USA

[2]Microelectronics-Photonics Program, University of Arkansas, Fayetteville, Arkansas 72701, USA

[3]Department of Physics, University of Arkansas, Fayetteville, Arkansas 72701, USA

[4]Institute for Nanoscience and Engineering, University of Arkansas, Fayetteville, Arkansas 72701, USA

[5]Department of Electrical Engineering and Physics, Wilkes University, Wilkes-Barre, Pennsylvania 18766, USA

[6]Thayer School of Engineering, Dartmouth College, Hanover, New Hampshire 03755, USA

[7]School of Electrical, Energy and Computer Engineering, Arizona State University, Tempe, Arizona 85287, USA

[8]Department of Engineering, University of Massachusetts Boston, Boston, Massachusetts 02125, USA

[9]Arktonics, LLC, 1339 South Pinnacle Drive, Fayetteville, Arkansas 72701, USA

†Corresponding author

**Abstract**

GeSn lasers enable monolithic integration of lasers on the Si platform using all-group-IV direct-bandgap materials. Although optically pumped GeSn lasers have made significant progress, the study of the electrically injected lasers has just begun only recently. In this work, we present explorative investigations of electrically injected GeSn heterostructure lasers with various layer thicknesses and material compositions. The cap layer total thickness was varied between 240 and 100 nm. At 10 K, a 240-nm-SiGeSn capped device had a threshold current density $J_{th} = 0.6$ kA/cm$^2$ compared to $J_{th} = 1.4$ kA/cm$^2$ of a device with 100-nm-SiGeSn cap due to an improved modal overlap with the GeSn gain region. Both devices had a maximum operating temperature $T_{max} = 100$ K. Device with cap layers of $Si_{0.03}Ge_{0.89}Sn_{0.08}$ and $Ge_{0.95}Sn_{0.05}$, respectively, were also compared. Due to less effective carrier (electron) confinement, the device with a 240-nm-GeSn cap had a higher threshold $J_{th} = 2.4$ kA/cm$^2$ and lower maximum operating temperature $T_{max} = 90$ K, compared to those of the 240-nm-SiGeSn capped device with $J_{th} = 0.6$ kA/cm$^2$ and $T_{max} = 100$ K. In the study of the active region material, the device with $Ge_{0.85}Sn_{0.15}$ active region had a 2.3× higher $J_{th}$ and 10 K lower $T_{max}$, compared to the device with $Ge_{0.89}Sn_{0.11}$ in its active region. This is likely due to higher defect density in $Ge_{0.85}Sn_{0.15}$ rather than an intrinsic issue. The longest lasing wavelength was measured as 2682 nm at 90 K. The investigations provide guidance to the future structure design of GeSn laser diodes to further improve the performance.

SiGeSn alloys have attracted considerable attentions in recent years as a versatile material system enabling all-group-IV-based optoelectronic device integration [1,2]. Specifically, direct bandgap GeSn offers a route towards monolithic integration of light sources on Si for mid-infrared applications [3]. The first set of GeSn lasers was demonstrated under optically pumping at temperatures up to 90 K [4]. Since then, the studies of GeSn laser have made inspiring leaps in maximum operating temperature ($T_{max}$) up to near-room-temperatures [5-7], with expanded wavelength coverage up to 4.6 μm [8], and with reduced thresholds of continuous wave operation [9]. The comparison between ridge waveguide lasers and micro-disk lasers provided further insight into the importance of optical confinement and heat dissipation [10]. Studies of the optically pumped lasers ranged from bulk lasers [11], to heterostructures [6], and multiple quantum well lasers [12,13]. The threshold was reduced by introducing carrier-confinement structures. Furthermore, an increase in the Sn composition of the GeSn active region led to lasing at elevated temperatures [6,7]. All these effects were studied under optical pumping.

The electrically injected GeSn laser diode is of even greater interest as they can be fully integrated on the Si-based group-IV photonics platform. However, to achieve lasing under electrical injection involves more challenges than that under optical pumping. For example, optical confinement in an optically pumped laser can be achieved by utilizing the low index Si substrate and the top air as cladding layers, but more challenging for electrically injected lasers because a metal contact is normally placed on the top surface of the device. Therefore, it is necessary to modify laser structure designs and to add more layers, which are difficult to grow. Another challenge is the increased free carrier absorption due to increased doping for more effective carrier injection through a pn junction, which increases laser threshold in comparison with optically pumped lasers, where carrier injection is easily achieved with the optical absorption of the pump light.

Such differences illustrate the need to design and evaluate the laser structures under electrical injection. Recently, an electrically pumped laser operational up to 100 K was demonstrated [14]. The devices had a minimum threshold at 0.6 kA/cm$^2$ and maximum peak power output of 2.7 mW/facet at 10 K. Compared to using n-type contact on the top surface, placing p-type contact on the top limited the hole leakage from the type-II band alignment between the cap layer and the active region. The thick active region and the low refractive index cap layer offered an optical confinement factor of 75%. Although this specific design leads to demonstration of electrically pumped lasers, still uncharted are the effects of layer thicknesses and alloy gain medium/barrier material selections, which offers fundamental guidance for future GeSn diode laser design.

In this work, several electrically injected GeSn heterostructure laser diodes with different cap layers and active layer material were compared. Increased cap thickness improves the optical confinement factor in the gain region and reduces the optical loss from the metal contact significantly. The devices with a 240-nm thick cap layer demonstrated reduced threshold, compared to the devices with a 100-nm-thick cap. Cap layer materials with different conduction band barrier heights were also studied. Because of improved electron confinement, the Si$_{0.03}$Ge$_{0.89}$Sn$_{0.08}$ capped devices with a 114 meV barrier exhibited a lower threshold and higher T$_{max}$, compared to the devices with Ge$_{0.95}$Sn$_{0.05}$ cap with a 58 meV barrier. Devices with 11% and 15% Sn composition in the GeSn active region are compared to probe the effect of intrinsic GeSn gain within the laser diode. Beyond the observation that lasing at a longer wavelength was recorded at 2682 nm at 90 K, the increase of Sn composition did not show improvement on threshold and T$_{max}$, implying that extrinsic strain-induced defect dislocations remain, causing the device performance to deteriorate.

EXPERIMENTS

**Layer structure of the laser diodes.** A schematic of the laser structure is described in Fig. 1(a). The five layers were epitaxially grown in the sequence of: (i) a 500 nm Ge buffer, n-type doped (phosphorus) at $1\times10^{19}$ cm$^{-3}$; (ii) a 700 nm spontaneous-relaxation-enhanced GeSn buffer, n-type doped at $1\times10^{19}$ cm$^{-3}$; (iii) a 1000 nm GeSn active region, undoped; (iv) the first GeSn or SiGeSn cap layer with p-type doping (boron) at $1\times10^{18}$ cm$^{-3}$; (v) a 50 nm (Si)GeSn cap layer with p-type doping at $1\times10^{19}$ cm$^{-3}$. The GeSn buffer layers had a varied Sn composition: 7-11% in samples A to D and 10-15% in sample E. The GeSn active region had a composition of 11% in samples A to D, and 15% in sample E.

**Experimental design.** Three sets of experiments were designed to study the GeSn laser structures, summarized in Fig. 1(b). The first set adjusts the total thickness of the cap layer in order to evaluate the optical mode-profile effect on the optical confinement factor as well as the absorption loss. Samples A and C have the total cap thickness of 240 nm compared to 100 nm in samples B and D, respectively. The second variable is the material used in the cap layers that changes conduction-band barrier height in the heterostructure. $Si_{0.03}Ge_{0.89}Sn_{0.08}$ with 114 meV of barrier height is used in samples A and B, compared to $Ge_{0.95}Sn_{0.05}$ with 58 meV barrier height used in samples C and D. The last experiment evaluates the device performance affected by the Sn composition in the GeSn active region. Sample E has a nominal 15% of Sn in the active region, compared to 11% in sample A. Note that sample A which has been reported in ref. 14 served as reference in this work. The key parameters are summarized in Table I.

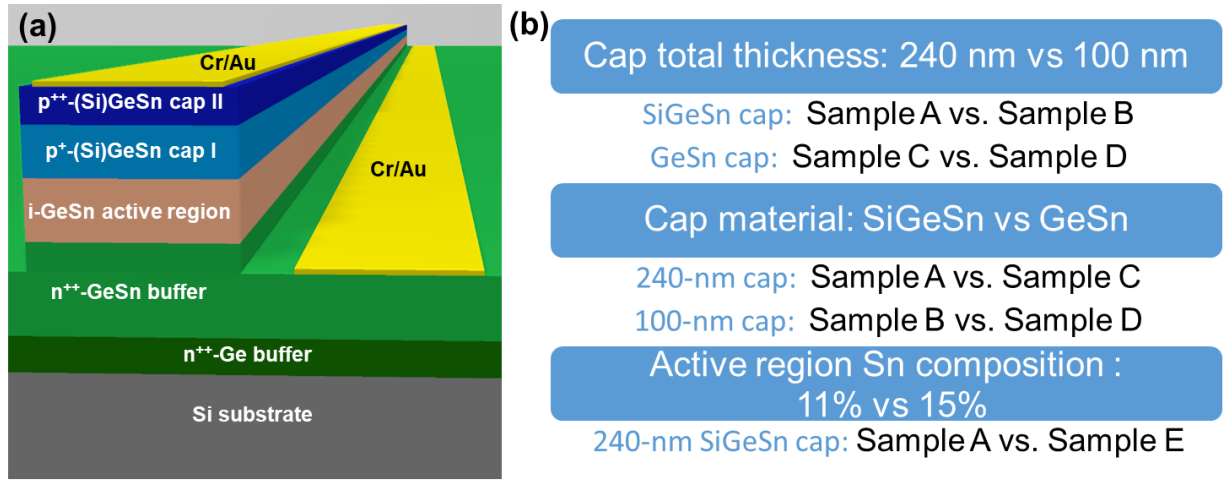

Fig. 1 (a) 3D schematic of the ridge waveguide laser structure; (b) three experiment groups are studied with tuning of the total cap thickness, cap layer material, and active region Sn composition.

RESULTS

**Observation of Lasing.**  To validate the lasing operation, the devices characteristics were measured using the same method that was detailed in ref. 14.  According to measurement results, all devices show unambiguous lasing characteristics under pulsed condition, as the threshold indicating the onset of lasing, dramatically reduced peak linewidth and increased peak intensity were clearly observed, which were acknowledged to identify the lasing of sample A in ref. 14.  The typical characterization results of sample E are shown in Fig. 2.  The detailed results of all other samples can be found in the supplementary.

Figure 2(a) displays the temperature dependent peak-power light output versus current injection (L-I) curves for sample E collected from a single facet.  The threshold characteristic can be clearly resolved.  The lasing was observed at temperatures 10 to 90 K, with the threshold from 1.4 to 3.6 kA/cm$^2$.  The characteristic temperature $T_0$ was extracted as 81 K.  The maximum output was measured of 0.7 mW/facet under 4 kA/cm$^2$ at 30 K.  From 10 to 50 K, the kink can be observed at ~ 1.5×$J_{th}$, which may be due to the lasing-mode switch.

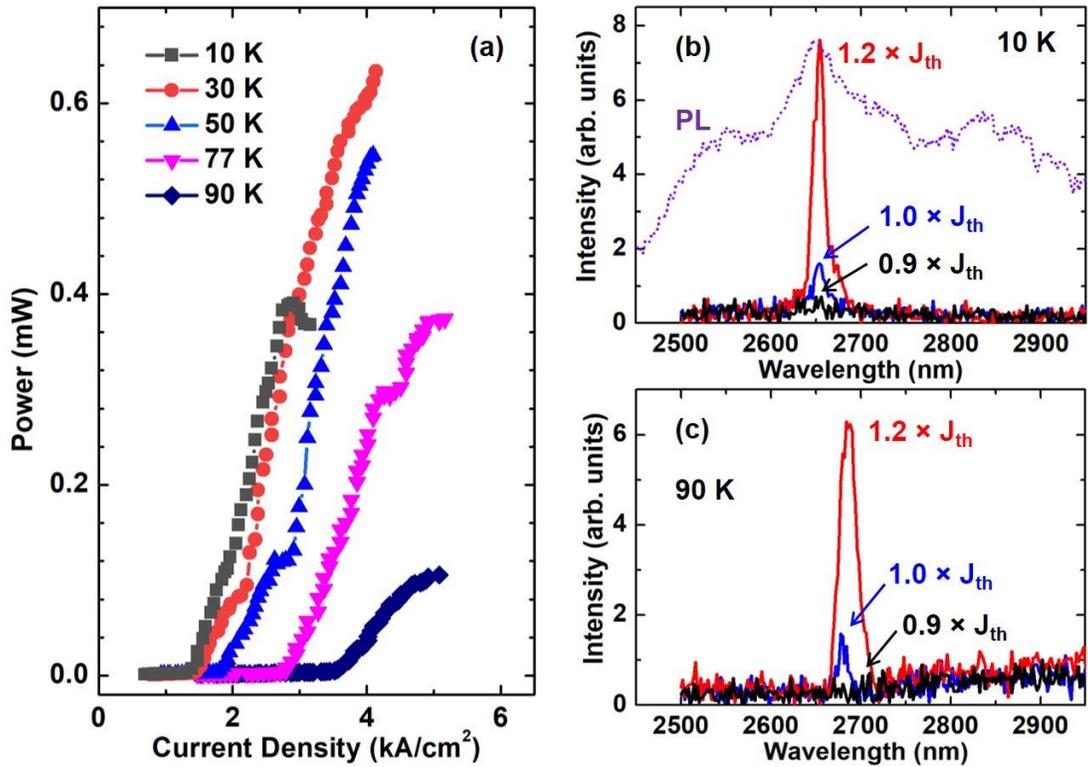

Fig. 2 Characterizations of Sample E. (a) Temperature dependent L-I curve; (b) Lasing spectra under injections below and above threshold at 10 K. The PL spectrum was also plotted for comparison; (c) Emission spectra under injections below and above threshold at 90 K.

Figure 2(b) shows the emission spectra below and above the lasing threshold at 10 K. The photoluminescence (PL) spectrum at 10 K was also plotted for comparison (dotted line). Compared to the PL peak, the significantly reduced emission peak linewidth indicates the onset of lasing. Note that based on our previous study [5, 7, 14], the devices feature multi-mode operation, and therefore the observed peak actually consists of multi-mode peaks (e.g. see the spectra in Fig. 2b for $1.2 \times J_{th}$ and that in Fig. 2c for $1.0 \times J_{th}$),, which cannot be further resolved due to the spectrometer resolution of 10 nm. Below threshold ($0.9 \times J_{th}$), a broad peak at ~2600 to 2720 nm with relatively weak intensity was obtained, suggesting the spontaneous emission. As current injection increases to above threshold, a narrower linewidth peak sitting on background spontaneous emission emerges, whose intensity dramatically increases at 1.2×$J_{th}$. This trend indicates the unambiguous lasing characteristic. The lasing peak wavelength was measured as 2654 nm at 1.2×$J_{th}$ at 10 K. The emission spectra at 90 K are plotted in Fig. 2(c). The spontaneous emission (at 0.9×$J_{th}$) from the active region cannot be resolved due to the low intensity. However, the lasing peak can be clearly identified at injection above the threshold. Under injection of 1.2×$J_{th}$ at 90 K, the lasing wavelength was measured as 2682 nm.

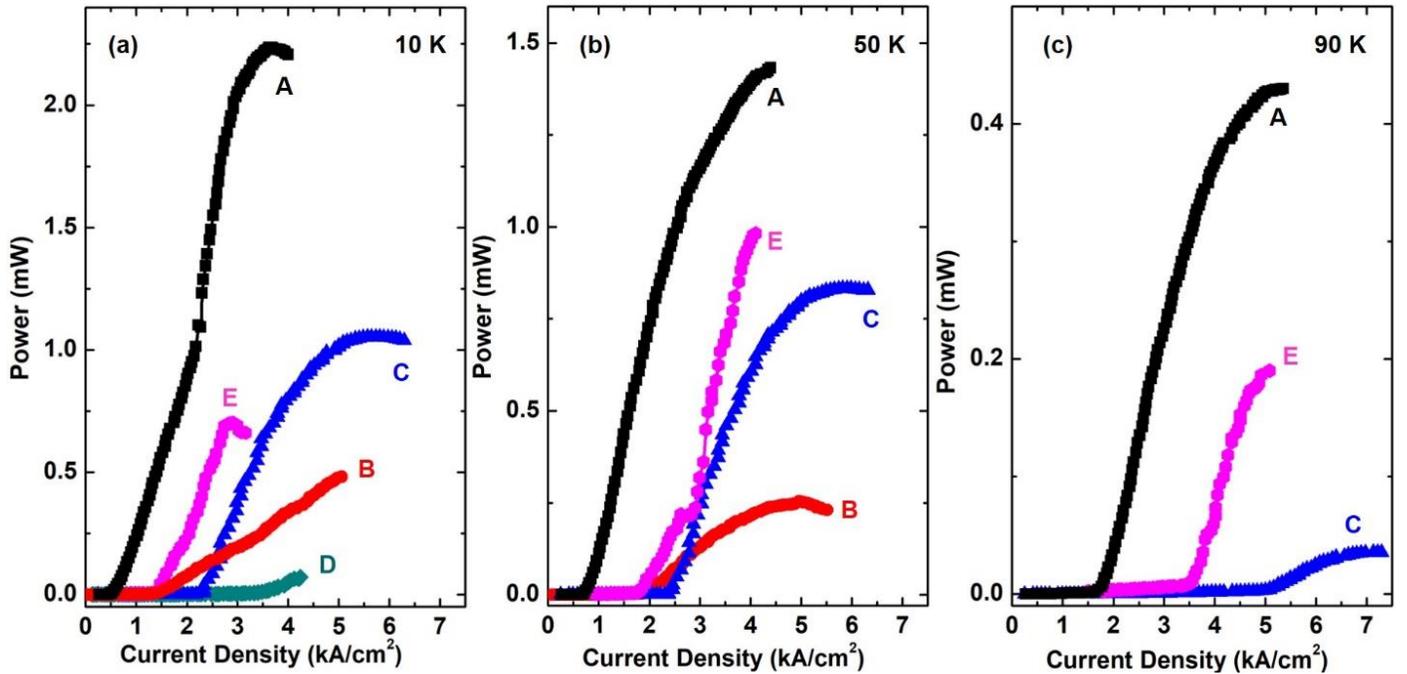

Fig. 3 L-I curves of each sample at (a) 10 K; (b) 50 K; and (c) 90 K.

**Light output-current injection characteristics.** The L-I curves of all devices at 10 K are plotted in Fig. 3(a). Sample A features the lowest threshold while sample D has the highest threshold, which are 0.6 and 3.4 kA/cm$^2$, respectively. It is worth noting that regarding the curve slope above the threshold, samples A, C, and E exhibit similar slope, which is higher than that of samples B and D. Except for reference sample A, sample C shows the maximum output power of 1.05 mW/facet under 5.5 kA/cm$^2$.

As temperature increases, sample D stops lasing above 10 K. Figure 3(b) shows L-I curves of all other samples at 50 K. The lowest and highest thresholds are 0.8 kA/cm$^2$ (sample A) and 2.5 kA/cm$^2$ (sample C), respectively. Sample B shows a lower curve slope above the threshold compared to the other three samples.

As temperature further increases, sample B stops lasing above 50 K. The L-I curves of samples A, C, and E at 90 K are shown in Fig. 3(c), with the corresponding thresholds of 1.8, 5.1, and 3.6 3.4 kA/cm$^2$, respectively. The maximum operational temperature of samples C and E is 90 K, while for sample A it is 100 K, which has been reported in ref. 14.

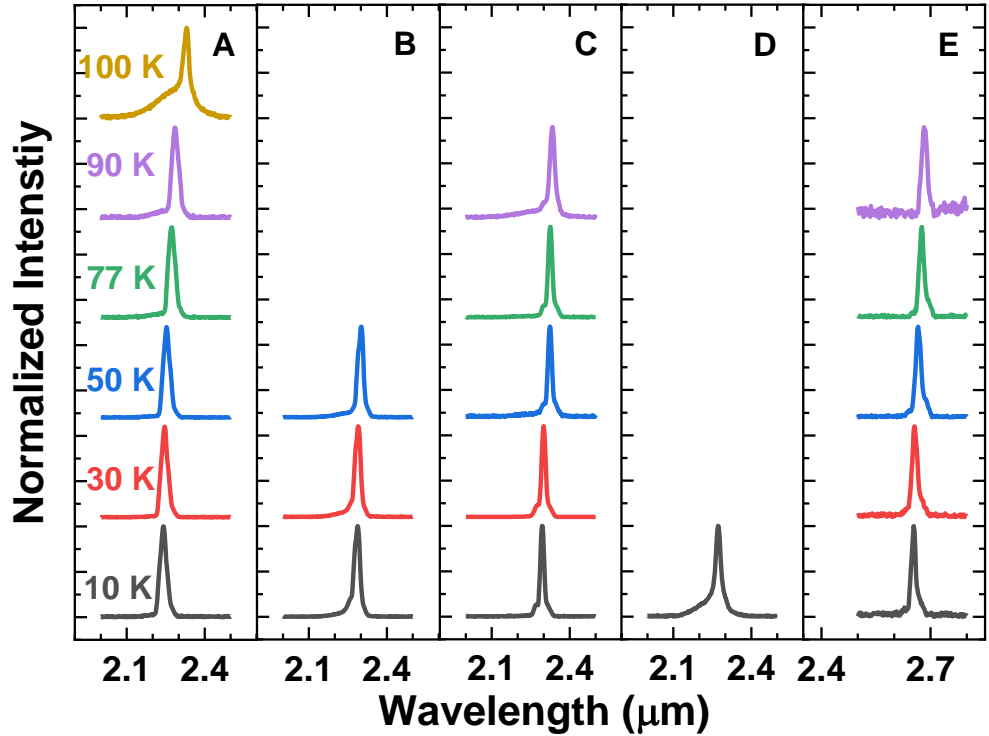

Fig. 4. Normalized spectra showing the lasing peaks under 1.1×J$_{th}$ injection for each sample at the corresponding operational temperatures.

Figure 4 summarizes the laser peak emission under injection of 1.1×J$_{th}$. At 10 K, the lasing peak of sample A was obtained at 2240 nm. While for samples B, C, and D the lasing peaks were observed at ~2270-2290 nm. This is due to the slight difference of Sn compositions in the active region (see the supplementary). The lasing peak of sample E is at 2654 nm, much longer than the rest of the samples due to the higher Sn composition (15% vs 11%) in active region. As the temperature increases, the peak emission shifts towards longer wavelength as expected, which represents a narrower bandgap at higher temperature. At 90 K, the lasing peak was obtained at 2682 nm, as shown in Fig. 4.

For samples A to E, the full widths at half maximum (FWHM) incorporating all lasing modes were measured as 32, 29, 18, 34, and 16 nm (under the 10 nm resolution spectrometer), respectively at 10 K. Our previous studies revealed that due to the relatively large area of cross section, all laser devices feature multi-mode operation. The high-resolution lasing spectra showing well-resolved multi-peaks of sample A was reported in ref. 14. The characteristics for all laser devices are summarized in Table I.

**Table I.** Summary of laser characteristics

| Sample | Cap Layer Material | Cap Layer Thickness (nm) | Sn % in Active Region | Threshold at 10 K (kA/cm$^2$) | Threshold at 77 K (kA/cm$^2$) | $T_{max}$ (K) | $T_0$ (0-$T_{max}$) (K) | Lasing Wavelength at 10 K (nm) |
|---|---|---|---|---|---|---|---|---|
| A | Si$_{0.03}$Ge$_{0.89}$Sn$_{0.08}$ | 240 | 11 | 0.6 | 1.4 | 100 | 76 | 2238 |
| B | Si$_{0.03}$Ge$_{0.89}$Sn$_{0.08}$ | 100 | 11 | 1.4 | N.A. | 50 | 119 | 2281 |
| C | Ge$_{0.95}$Sn$_{0.05}$ | 240 | 11 | 2.4 | 3.1 | 90 | 123 | 2294 |
| D | Ge$_{0.95}$Sn$_{0.05}$ | 100 | 11 | 3.4 | N.A. | 10 | N.A. | 2272 |
| E | Si$_{0.03}$Ge$_{0.89}$Sn$_{0.08}$ | 240 | 15 | 1.4 | 2.9 | 90 | 81 | 2654 |

DISCUSSION

**Thickness of cap layer (first experimental group in Fig. 1b).** The total thickness of the cap layer affects considerably the lasing performance. Both comparisons (A vs B and C vs D: 240 nm vs 100 nm) show the same trend, no matter whether SiGeSn or GeSn is used as cap layer material - that a thicker cap device has a lower lasing threshold as well as higher maximum operating temperature: at 10 K; sample A has a threshold that is 0.43× of that of sample B; while sample C has a lower threshold that is 0.70× of sample D. For the maximum operating temperature, 100 K for sample A vs 50 K for sample B, and 90 K for sample C vs 10 K for sample D were observed. The change of cap layer thickness affects the laser performance by the following factors: (i) optical loss from the metal contact plays the major role, (ii) free carrier absorption within the heavily doped cap layers, and (iii) optical confinement factor in the active region.

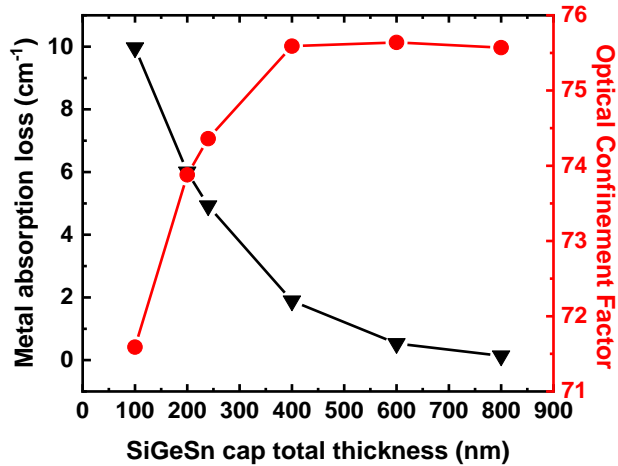

Fig. 5. Calculated metal absorption loss and optical confinement factor as functions of SiGeSn cap layer total thickness.

The optical loss on the metal contacts consists of two origins: absorption loss in the metal layer and scattering loss due to the surface roughness. The metal absorption loss is plotted as a function of SiGeSn cap total thickness in Fig. 5. As cap layer thickness increases, the absorption loss decreases. The metal absorption loss is halved by increasing the cap thickness from 100 (B and D) to 240 nm (A, C, and E) and is below 1 cm$^{-1}$ when the cap thickness reaches beyond 600 nm. On the other hand, the increase of cap layer thickness decreases the modal overlap with the metal layer, and therefore the scattering loss due to the metal surface roughness is reduced. The calculated metal absorption loss is summarized in Table II.

For each sample, the free carrier absorption loss of each layer was calculated based on the doping level and the carrier injection [15], as summarized in Table II. The active region is assumed to have carrier density at $n = p = 5\times10^{18}$ cm$^{-3}$. The n-type doped GeSn buffer shows almost identical values (0.17 cm$^{-1}$), while the p-type doped cap layer exhibits lower values (~1.0 cm$^{-1}$) for samples with thicker cap and higher values (2.3 and 2.7 cm$^{-1}$) for samples with thinner cap. The active region makes major contribution to the overall free carrier absorption loss (>20.0 cm$^{-1}$), which is a result of thick layer and high confinement factor.

**Table II.** Calculated Loss

| Sample # | P-type Cap Free Carrier Absorption Loss (cm$^{-1}$) | Active Region Free Carrier Absorption Loss (cm$^{-1}$) | N-type Buffer Free Carrier Absorption Loss (cm$^{-1}$) | Metal Absorption Loss (cm$^{-1}$) | Total Loss (cm$^{-1}$) |
|---|---|---|---|---|---|
| A | 0.7 | 22.3 | 0.17 | 4.9 | 28.1 |
| B | 2.3 | 22.9 | 0.15 | 10 | 35.4 |
| C | 0.8 | 22.4 | 0.17 | 4.9 | 28.3 |
| D | 2.7 | 23.0 | 0.15 | 10 | 35.9 |
| E | 1.2 | 23.1 | 0.17 | 5.3 | 29.8 |

The optical confinement factor in the active region is calculated and shown in Fig. 5. Due to the thick Ge and GeSn buffer of ~1200 nm, the thicker cap is designed to increase the overlap between optical mode and the active region. However, with a relatively thick active region (~1000 nm), as the cap layer thickness increases from 100 to 800 nm, the optical confinement factor ranges from 71.5% to 75.5%, showing a non-significant change. In this work, the optical confinement factors of 240-nm-cap samples and 100-nm-cap samples are 74.0% and 71.5%, respectively.

Overall, it can be seen that the free carrier absorption loss and optical confinement show less sensitivity to the cap layer thickness, while the metal absorption loss dramatically decreases as the cap layer thickness increases. Note that the total loss in Table II is underestimated as the metal scattering loss was not included.

**Cap layer materials (second experimental group in Fig. 1b).** The cap layer also serves as the top barrier with respect to active region. $Si_{0.03}Ge_{0.89}Sn_{0.08}$ and $Ge_{0.95}Sn_{0.05}$ are compared as the cap layer materials. Note that both cap materials feature tensile strain due to relatively smaller lattice constants with respect to the active layer, leading to type-II band alignment being obtained in the light hole (LH) in the valence band (VB) at the cap/active interface [14], which creates a hole leakage channel. This issue was addressed by injecting holes from the top so that the holes flow from the p-type cap layer towards the n-type GeSn buffer, where the barrier at the active/GeSn buffer interface could confine the holes in the active region, as described in ref. 14. Therefore, the electron confinement offered by cap layers is a focus in this experiment.

The conduction band (CB) barrier height of each sample was calculated. For both comparisons (A vs C and B vs D: $Si_{0.03}Ge_{0.89}Sn_{0.08}$ vs $Ge_{0.95}Sn_{0.05}$), no matter whether the cap layer thickness is 240 nm or 100 nm, the samples using $Si_{0.03}Ge_{0.89}Sn_{0.08}$ cap have the barrier height of 114 meV, while the samples using $Ge_{0.95}Sn_{0.05}$ cap show a lower barrier height of 58 meV. The higher barrier height improves the device performance due to better electron confinement: (i) at 10 K, sample A has a much lower threshold that is 0.25× of that of sample C, while sample B has a threshold that is 0.41× of sample D. The similar threshold reduction was also reported on III-V laser devices [16]; (ii) for the maximum operating temperature, 100 K for sample A vs 90 K for sample C, and 50 K for sample B vs 10 K for sample D were obtained. As temperature increases, due to increased thermal energy $k_BT$ that can be absorbed by carriers, the electron confinement capability is weakened, leading to the result that the samples using $Ge_{0.95}Sn_{0.05}$ cap (lower barrier height) stop lasing prior to those using $Si_{0.03}Ge_{0.89}Sn_{0.08}$ cap (higher barrier height). Note for sample D, because of its thin $Ge_{0.95}Sn_{0.05}$ cap (100 nm), it can only lase at 10 K.

The CB barrier height between the GeSn active and the GeSn buffer was calculated for each sample as well. Samples A to D show similar hole barrier height of ~30 meV. At room temperature, such a barrier is slightly greater than 1 $k_BT$, and therefore it is insufficient for hole confinement. To improve the hole confinement, inserting a wider bandgap SiGeSn (with appropriate Si and Sn compositions) as the tom barrier can be considered.

**Sn compositions in the active region.** Samples A ($Ge_{0.89}Sn_{0.11}$) and E ($Ge_{0.85}Sn_{0.15}$) are compared for the active region materials. Our previous studies on GeSn optically pumped lasers indicated that the increase of Sn composition in the active region led to the increase of the maximum operating temperature and reduction of lasing threshold [5-7]. This is attributed to the higher-Sn-induced greater directness of bandgap in the active region, which facilitates the electron populating the Γ valley. However, in this experiment, the electrically injected device with higher Sn composition neither shows reduced threshold nor exhibits increased operating temperature: At 10 and 77 K, the thresholds of sample E (1.4 and 2.9 kA/cm$^2$) are doubled compared to those of sample A (0.6 and 1.4 kA/cm$^2$); the maximum lasing temperature of 90 K for sample E is 10 K lower than that of sample A. This can be interpreted as follows: (i) to achieve population inversion, the optically pumped lasers rely on optical absorption. For a certain wavelength of incident light, the absorption coefficient increases as Sn composition in the GeSn active region increases due to narrower bandgap. However, for electrically injected lasers, the carrier injection efficiency shows insensitivity to Sn composition in the active region, and therefore there is no enhanced absorption associated with the higher Sn device; (ii) according to current material growth properties, a higher Sn composition layer may feature a high defect density, which deteriorates the lasing performance. Recent studies on GeSn DHS LEDs show similar results; that an increase of Sn composition does not directly improve the emission intensity [17, 18; (iii) the GeSn active layer with higher Sn composition has a larger lattice constant, which leads to larger lattice mismatch with the cap layer, resulting in increased defect density at the active/cap interface. These defects act as recombination centers, and thus reduce the carrier injection efficiency. Therefore, for the device with higher Sn in the active region, the cap layer needs to be selected to minimize the interface defect density. The SiGeSn with appropriate Si and Sn compositions is a suitable candidate. Current material growth shows limited availability of Si and Sn compositions, which, however, can be improved by the advance of future growth techniques; (iv) Sample A has a slightly longer cavity length of 1.7 mm compared to 1.3 mm for sample E. The longer cavity features reduced mirror loss, leading to a better performance. Note that the difference of mirror loss between samples A and E is relatively small (5.5 cm$^{-1}$ vs 7.2 cm$^{-1}$), and thus it is not a dominant factor in this experiment.

**Additional considerations.** In addition to the loss mechanism abovementioned, other factors that affect the device performance are elaborated as follows: (i) active layer thickness. In this work, the design of sample structure is inherited from the former optically pumped lasers, in which a thicker active layer is preferred, aiming to have higher light absorption. However, such absorption enhancement with a thicker active layer does not apply to electrically injected lasers. In fact, growing a thinner active region could effectively reduce the threshold [19]. (ii) absorption loss in the GeSn buffer and the SiGeSn/GeSn cap. Since the GeSn buffer is almost relaxed and heavily doped, the high density of dislocations due to lattice mismatch would create defect energy levels in the band gap, resulting in additional absorption loss. The absorption tails have been reported in GeSn materials, which lead to below-band-gap absorption with the orders of 1~10 cm$^{-1}$ for extra loss [20, 21]. Likewise, once the thicker cap layer beyond the critical thickness is employed, the gradually relaxed material would introduce considerable density of dislocations. Therefore, the lattice-matched cap layer is desirable, which again relies on the advance of SiGeSn material growth capability; (iii) the scattering loss induced by sidewall roughness. The relatively rough sidewall would result in additional scattering loss, especially for the device with a relatively wide ridge. Reducing the sidewall roughness by improving the fabrication procedure is a viable solution to reduce the scattering loss.

In conclusion, electrically injected GeSn laser diodes were studied with an evaluation of three factors in the structure: cap layer thickness, cap materials, and active region materials. The 240-nm capped devices feature lower metal absorption loss, offering a reduction of threshold and an elevation of maximum operating temperature compared to 100-nm capped devices; The devices with a $Si_{0.03}Ge_{0.89}Sn_{0.08}$ cap show lower threshold and higher maximum operating temperature compared to the devices using a $Ge_{0.95}Sn_{0.05}$ cap, which is due to the higher barrier height at the active/cap interface when employing the $Si_{0.03}Ge_{0.89}Sn_{0.08}$ cap; The 15% Sn in the active region, however, does not improve the device performance in terms of lasing threshold and temperature compared to an 11% Sn device, when the same $Si_{0.03}Ge_{0.89}Sn_{0.08}$ cap was used. The performance of high-Sn composition devices can be improved by employing a thinner active layer and a cap with less lattice mismatch, which could reduce the dislocation density. The maximum lasing peak wavelength was measured at 2682 nm for $Ge_{0.85}Sn_{0.15}$ devices at 90 K.

METHODS

**Sample growth.** The sample structures were grown on the 200-mm Si (100) wafer using reduced pressure chemical vapor deposition. The heterostructure was grown on the Si wafer starting with a 500-nm Ge buffer. The Ge buffer was in-situ annealed to limit the defects. The 700-nm GeSn buffer was grown with the recipe of nominal 11% Sn for samples A-D and 15% Sn for sample E. The dislocations induced by the crystalline strain relaxation is expected within the GeSn buffer layer. And the relaxation leads to a Sn gradient within the GeSn buffer. The dislocations are expected to cause a Sn gradient within the GeSn buffer and confined within some few hundreds of nanometers. The GeSn active region was then grown with the same recipe and low defect density. The standard in-situ dopant gas was introduced to form the n-type and p-type regions.

**The photoluminescence (PL) measurement.** The PL was measured from the cap-removed sample. The heterostructure sample was chemically etched. And the top 600 nm of structure was removed, and the active region was exposed at the surface to be characterized. The PL emission was excited by using the 532-nm continuous wave laser at 500 mW. The emission was collected through the spectrometer with the InSb detector. The signal was recorded via a standard lock-in technique.

**Device fabrication.** The structures were fabricated into ridge waveguide laser diodes. The 80-μm wide ridge structure was formed by standard photolithography and wet chemical etching. The ridge height is 1.4 μm in order to reach the n-type GeSn buffer layer. The metal contacts (10 nm Cr + 350 nm Au) were deposited by using an electron beam evaporator. The Si substrate on the back side was reduced to ~140 μm thickness by lapping, followed by a cleaving to form the facets. The laser diodes were wire-bonded onto a Si carrier with isolated Au bonding pads.

**Electrically injected laser measurements.** The chip was mounted in a temperature-controlled cryostat for the characterization. The device operated under a pulsed mode to avoid Joule heating. The pulsed voltage source was used as the pumping source with a 1-kHz repetition rate and 700 ns of pulse width. The emission power and spectra were characterized using a grating-based spectrometer equipped with a liquid-nitrogen-cooled InSb detector (detection range 1.0-5.5 μm). The spectra presented in this paper focus on comparing the peak wavelength and on demonstrating the emission below and above the lasing threshold. Therefore, the resolution of the spectrometer is set at 10 nm, which is compromised to obtain a reasonable signal-to-noise ratio for the spectrum, especially at the low intensity level below the threshold. The absolute optical power was measured using a calibrated power meter. The emission from single facet was coupled by a pair of convex lenses and focused on the power meter. Note that the reflection and absorption losses through the cryostat window and lens were calculated and were added to the power meter readout aiming to report more accurate emission power. The emission was then guided onto the InSb detector through the spectrometer for the L-I and spectral measurements. The detailed description of the high-resolution spectra and the absolute power calibration process can be found in Ref. 14.

ASSOCIATED CONTENT

**Supporting information**

Supplementary information is available in the online version of the paper. Reprints and permissions information are available online at www.nature.com/reprints. Correspondence and requests for materials should be addressed to S.Y.

AUTHOR INFORMATION

**Corresponding author**

**Shui-Qing Yu** – Department of Electrical Engineering, University of Arkansas, Fayetteville, Arkansas 72701, USA

Institute for Nanoscience and Engineering, University of Arkansas, Fayetteville, Arkansas 72701, USA

Email: syu@uark.edu

**Notes**

The authors declare no competing financial interests.

ACKNOWLEDGEMENTS

The authors acknowledge the financial support from the Air Force Office of Scientific Research (AFOSR) (Grant Nos. FA9550-18-1-0045, FA9550-19-1-0341). Dr. Wei Du appreciates support from Provost's Research & Scholarship Fund at Wilkes University.